\documentclass[conference]{IEEEtran} 
\IEEEoverridecommandlockouts
\usepackage{cite}
\usepackage{amsmath,amssymb,amsfonts}
\usepackage{algorithmic}
\usepackage{graphicx}
\usepackage{textcomp}
\usepackage{xcolor}
\usepackage{url}
\def\BibTeX{{\rm B\kern-.05em{\sc i\kern-.025em b}\kern-.08em
    T\kern-.1667em\lower.7ex\hbox{E}\kern-.125emX}}
\begin{document}

\title{DBOS Network Sensing: A Web Services Approach to Collaborative Awareness
\thanks{
Research was sponsored by the Department of the Air Force Artificial Intelligence Accelerator and was accomplished under Cooperative Agreement Number FA8750-19-2-1000. The views and conclusions contained in this document are those of the authors and should not be interpreted as representing the official policies, either expressed or implied, of the Department of the Air Force or the U.S. Government. The U.S. Government is authorized to reproduce and distribute reprints for Government purposes notwithstanding any copyright notation herein.
Use of this work is controlled by the human-to-human license listed in Exhibit 3 of https://doi.org/10.48550/arXiv.2306.09267
}
}

\author{\IEEEauthorblockN{
Sophia Lockton$^1$, Jeremy Kepner$^1$, Michael Stonebraker$^{1,2}$, Hayden Jananthan$^1$, \\  LaToya Anderson$^1$, William Arcand$^1$, David Bestor$^1$, William Bergeron$^1$, Alex Bonn$^1$, Daniel Burrill$^1$, \\ Chansup Byun$^1$, Timothy Davis$^3$, Vijay Gadepally$^1$, Michael Houle$^1$, Matthew Hubbell$^1$, Michael Jones$^1$, \\ Piotr Luszczek$^{1,4}$, Peter Michaleas$^1$, Lauren Milechin$^1$, Chasen Milner$^1$, Guillermo Morales$^1$, Julie Mullen$^1$, \\ Michel Pelletier$^5$, Alex Poliakov$^2$, Andrew Prout$^1$,  Albert Reuther$^1$, Antonio Rosa$^1$, Charles Yee$^1$, Alex Pentland$^1$ 
\\
\IEEEauthorblockA{
$^1$MIT, $^2$DBOS, $^3$Texas A\&M, $^4$University of Tennessee, $^5$OneSparse
}}}
\maketitle

\begin{abstract}
DBOS (DataBase Operating System) is a novel capability that integrates web services, operating system functions, and database features to significantly reduce web-deployment effort while increasing resilience.  Integration of high performance network sensing enables DBOS web services to collaboratively create a shared awareness of their network environments to enhance their collective resilience and security.   Network sensing is added to DBOS using  GraphBLAS hypersparse traffic matrices via two approaches: (1) Python-GraphBLAS and (2) OneSparse PostgreSQL.  These capabilities are demonstrated using the workflow and analytics from the IEEE/MIT/Amazon Anonymized Network Sensing Graph Challenge.  The system was parallelized using pPython and benchmarked using 64 compute nodes on the MIT SuperCloud.  The web request rate sustained by a single DBOS instance was ${>}10^5$, well above the required maximum, indicating that network sensing can be added to DBOS with negligible overhead.  For collaborative awareness, many DBOS instances were connected to a single DBOS aggregator.  The  Python-GraphBLAS and OneSparse PostgreSQL implementations scaled linearly up to 64 and 32 nodes respectively.  These results suggest that DBOS collaborative network awareness can be achieved with a negligible increase in computing resources.
\end{abstract}


\section{Introduction}

Modern web applications, particularly those deployed in distributed cloud environments, face increasingly sophisticated and coordinated cyber threats \cite{kwon2019survey}. Traditional intrusion detection systems (IDSs), which rely on signature-based or anomaly-based techniques, are typically deployed in isolation \cite{kwon2019survey, locasto2005towards}. Collaborative awareness offers a paradigm shift: rather than monitoring in isolation, systems share anonymized summaries of observed traffic to construct a shared, real-time view of network activity.  \begin{figure}[htbp]
\centering
\includegraphics[width=1.0\columnwidth]{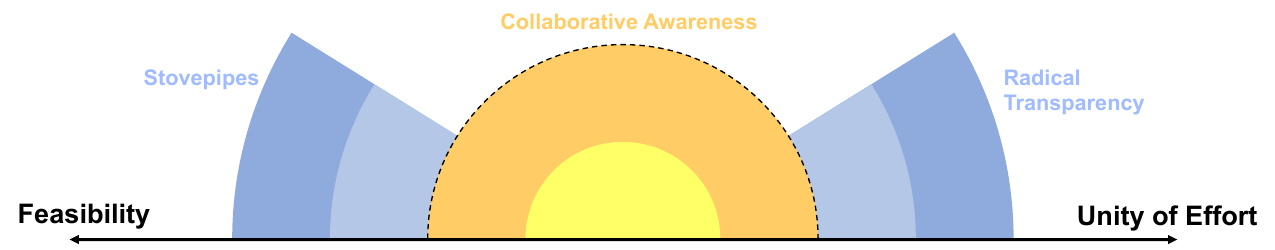}
\caption{\textbf{Collaborative Awareness}. This approach enhances both the efficiency and precision of threat detection, allowing systems to identify attacks that would likely go unnoticed by isolated monitoring setups \cite{badsha2019privacy, locasto2005towards, katti2005collaborating}. Stovepipe isolation (left) prevents lateral communication between organizational silos, maximizing privacy but limiting visibility. Radical transparency (right) mandates universal disclosure of all security-relevant data, theoretically maximizing defensive capabilities but proving infeasible due to data sharing restrictions. Collaborative awareness (center) enables exchange of sufficient  data, providing actionable intelligence while maintaining the highest regard for privacy.}
\label{fig:CollaborativeAwareness}
\end{figure}
Figure~\ref{fig:CollaborativeAwareness} illustrates the positioning of collaborative awareness between stovepiped isolation and radical transparency that enables systems to exchange actionable intelligence while maintaining privacy and regulatory compliance \cite{katti2005collaborating, atkins2021cooperation, demchak2021achieving, atkins2022web, kepner2021zero, weed2024beyond, atkins2025cyberspace}.

Our implementation operationalizes collaborative awareness by adapting the workflow defined in the Anonymized Network Sensing Graph Challenge, which formalizes network analysis through the construction, aggregation, and analysis of anonymized traffic matrices \cite{jananthan2024anonymized}. This approach is enabled by the GraphBLAS which represents network traffic as anonymized source/destination IP (Internet Protocol) pairs in hypersparse matrices. With the efficient hypersparse linear algebra capabilities of GraphBLAS, this compact yet information rich format enables real-time aggregation and analysis of network events on large spatial and temporal scales with commodity hardware \cite{pisharody2021realizing}.

\begin{figure}[htbp]
\centering
\includegraphics[width=0.9\columnwidth]{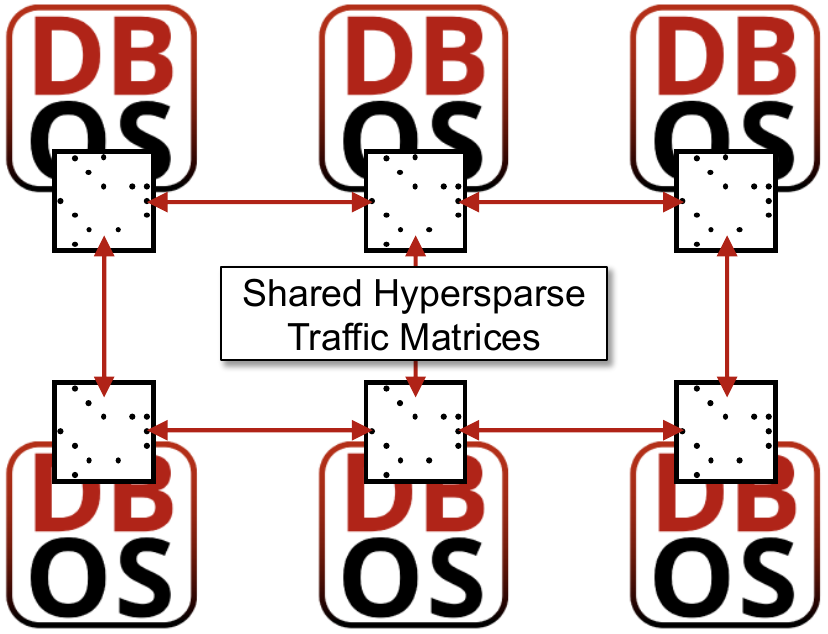}
\caption{\textbf{DBOS Collaborative Awareness Architecture.} Integration of GraphBLAS hypersparse traffic matrix construction into DBOS enables  DBOS web services to collaboratively create a shared awareness of their network environments to enhance their collective resilience and security.}
\label{fig:DBOS-SharedMatrices}
\end{figure}

We embed this framework directly into DBOS (Database-Oriented Operating System), a distributed runtime where all operating system services are implemented as transactional stored procedures on tables in a PostgreSQL relational database \cite{dbos_web, li2022progress}. The DBOS execution library simplifies the development of transactional web applications. Centralizing the storage of all web service activity into a database provides an ideal environment for embedding real-time monitoring and analytics \cite{skiadopoulos2021dbos}. By embedding GraphBLAS traffic matrices directly into its runtime fabric, DBOS instances are able to summarize local network activity, share traffic aggregates, and construct a global view of network activity in real-time (see Figure~\ref{fig:DBOS-SharedMatrices}). This built-in support for collaborative awareness means that developers building web services on DBOS receive state-of-the-art security as a byproduct of deployment.

This paper introduces Python and PostgreSQL-based implementations of collaborative awareness workflows within DBOS with global coordination using a shared file system or message passing interface. We evaluate the performance of three strategies in a high-throughput, multi-node setting, demonstrating that our system comfortably meets the demands of modern web services and scales efficiently with the number of participating instances. This work ultimately offers a practical and economically viable model for embedding collaborative security into everyday web infrastructure.

\section{Background}

Collaborative Intrusion Detection Systems (CIDS) enable shared observability across multiple nodes to effectively identify cyber security threats. CIDS architectures include: centralized systems (with a central analysis node), decentralized systems (hierarchical aggregation), and fully distributed systems (peer-to-peer collaboration). Each model presents trade-offs in detection accuracy, scalability, and complexity. Centralized and hierarchical systems face bottlenecks and single points of failure, while distributed systems introduce coordination challenges and trust management overhead \cite{vasilomanolakis2015taxonomy, sharma2020evaluation, bougueroua2021survey, nasir2024securing}.

The emergence of hypersparse traffic matrices as an extremely compact representation of network activity has mitigated many of the traditional concerns associated with centralized models. The dimensions of the matrix are equivalent to the full range of IPv4 addresses with each row corresponding to a source IP and each column to a destination IP. Recent work has demonstrated that anonymized hypersparse traffic matrices can be constructed with a standard server at rates exceeding 50 million packets per second, comparable to the capacity of a typical 400 Gigabit network link \cite{jones2022graphblas, houle2024hypersparse, han2024extracting}. The Anonymized Network Sensing Graph Challenge framework formalizes this approach \cite{jananthan2024anonymized}, encoding live traffic as hypersparse GraphBLAS matrices and preserving structural semantics through prefix-preserving anonymization schemes like CryptoPAN \cite{xu2002prefix}. This avoids exposing sensitive network event data while retaining analytic utility \cite{kepner2014achieving, jananthan2024anonymized}. It is important to note that a complete privacy-preserving scheme also requires formal organizational controls through data sharing agreements. These contractual mechanisms complement technical anonymization by providing legal enforcement against misuse, ensuring that traffic data can be shared across organizations while minimizing privacy risks \cite{jananthan2024anonymized, kepner2021zero, caida_aua}.

The SuiteSparse GraphBLAS library, an implementation of the GraphBLAS API standard, directly supports hypersparse matrices \cite{davis2019algorithm}. The SuiteSparse API is exposed in Python by the Python-GraphBLAS library and in PostgreSQL via the OneSparse extension \cite{python-graphblas, onesparse}. OneSparse is currently the only sparse graph implementation for a SQL database and exposes GraphBLAS types and methods natively in PostgreSQL. For this implementation, OneSparse was added as an extension to a custom PostgreSQL installation. The hypersparse GraphBLAS representation leveraged in both the Python and PostgreSQL implementations is powerful enough to enable efficient statistical analysis on extensive network data sets with modest computational overhead \cite{kawaminami2022large, jones2023deployment, jones2022graphblas}. 

Traffic matrix-based monitoring enables both statistical and machine learning approaches to anomaly detection. Common network quantities readily computed from traffic matrices can often be modeled as power-law distributions \cite{kepner2024normal}. Statistical deviations from these distributions are easily identifiable and interpretable indicators of anomalous behavior \cite{kepner2021zero}. Streams of traffic matrices can also be treated as image-like tensors, allowing deep learning models to learn accurate background models of network behavior for identifying anomalies \cite{kepner2021zero, soule2004identify, zhang2005estimating, tune2013internet}.

To guide the development of such analytics pipelines, the Anonymized Network Sensing Graph Challenge defines a standardized workflow for constructing, anonymizing, and analyzing traffic matrices at scale. It defines a reference pipeline (see Figure 4 in \cite{jananthan2024anonymized}) for processing raw packet data into traffic matrices and nine network quantities to be computed from the traffic matrices which in this context are: number of valid web requests ($N_v$), unique links, unique sources, max source requests, max source fan-out, unique destinations, max destination requests, and max destination fan-in (see Table I in \cite{jananthan2024anonymized}). The following implementation extends the Anonymized Network Sensing Graph Challenge framework to the web services layer by deriving traffic matrices from the sources and destinations in DBOS web requests.  The entire pipeline is implemented with the exception of anonymization which is deferred to allow a more in-depth integration into the DBOS authentication/authorization architecture.

\section{Workflow Design}
\begin{figure}[htbp]
\centering
\includegraphics[width=1.0\columnwidth]{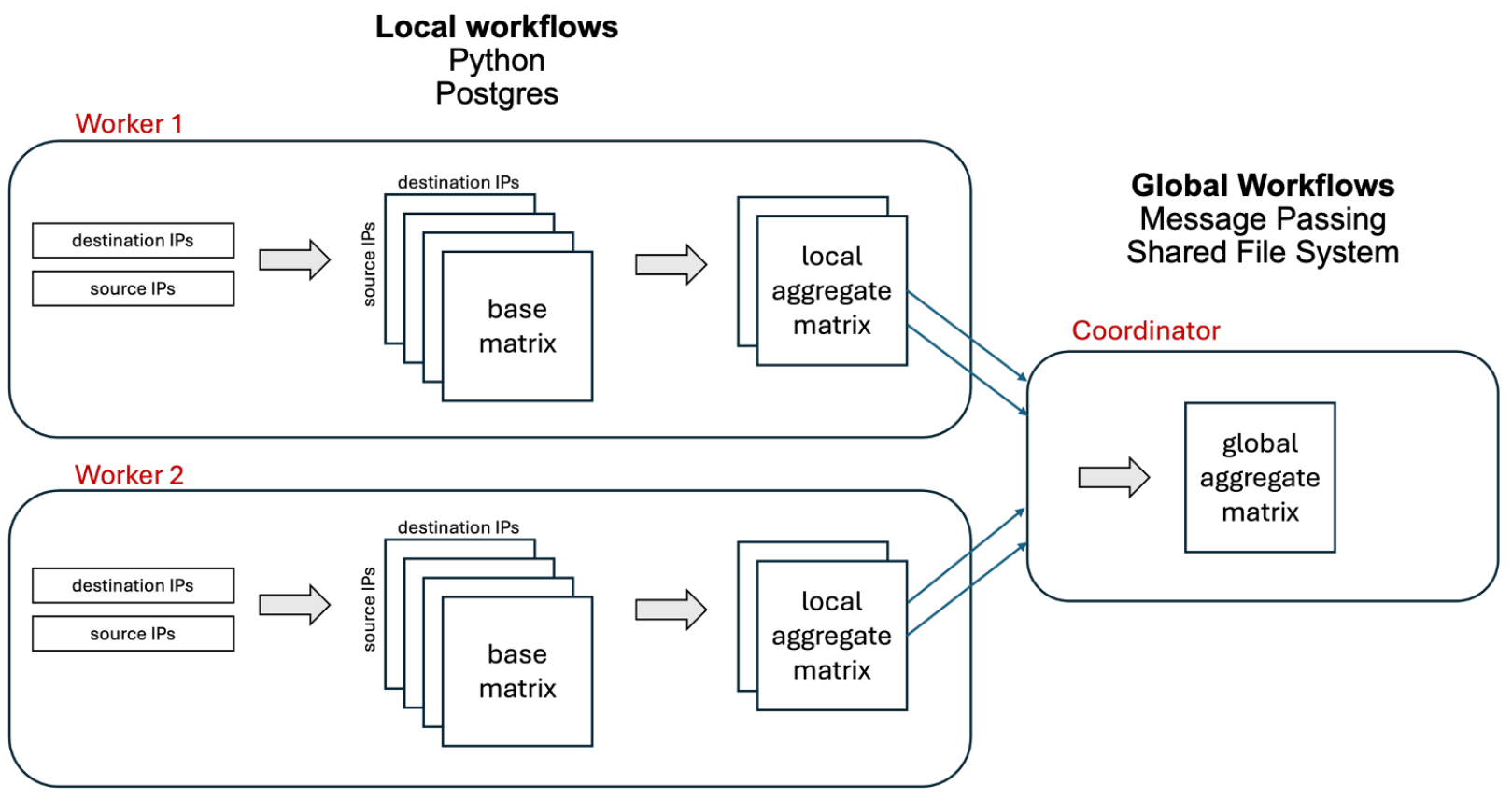}
\caption{\footnotesize \textbf{Overview of Local and Global Workflows.} Each DBOS instance constructs base GraphBLAS traffic matrices from observed source and destination IP pairs. These are aggregated into a local GraphBLAS traffic matrix, which is then transmitted to a coordinator. The coordinator aggregates the local traffic matrices into a global GraphBLAS traffic matrix, producing a system-wide summary of network activity.}
\label{fig:WorkflowOverview}
\end{figure}

The implementation is divided into two main phases: local traffic matrix aggregation and global traffic matrix aggregation. Figure~\ref{fig:WorkflowOverview} provides a broad overview of the different types of local and global workflows as well as the common steps across the different workflows.
\subsection{Local Aggregation (Python)}
\begin{figure}[htbp]
\centering
\includegraphics[width=1.0\columnwidth]{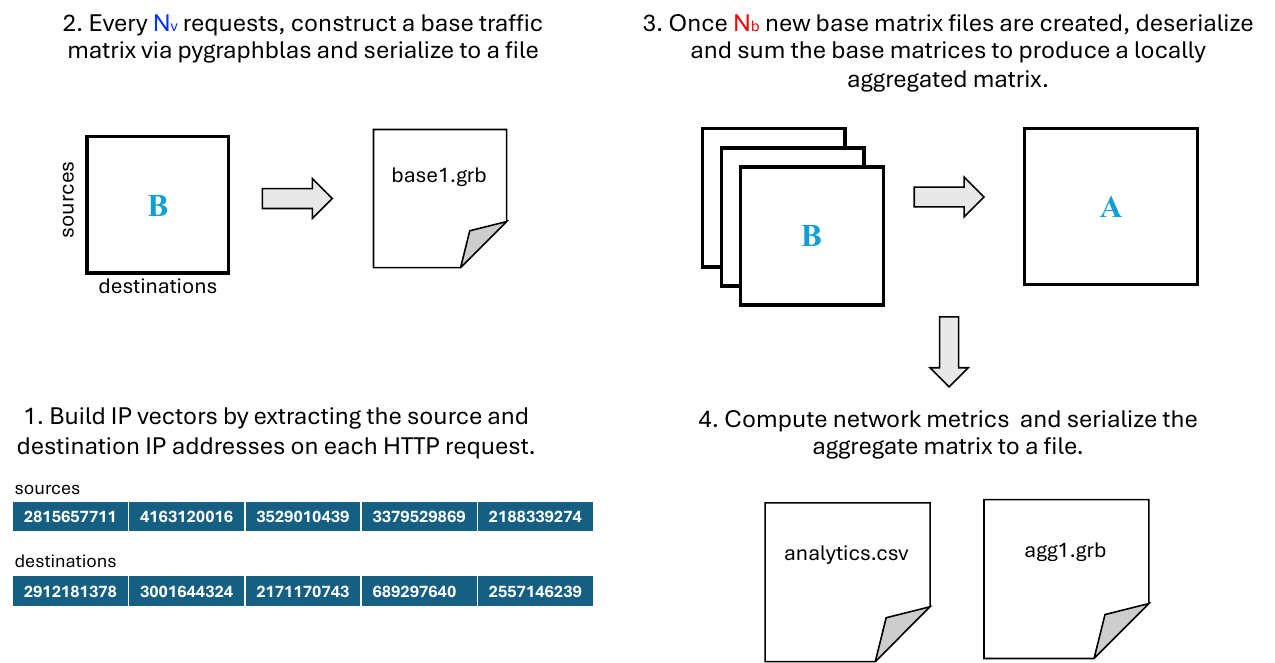}
\caption{\textbf{Python Local Aggregation Workflow}}
\label{fig:local_python}
\end{figure}

In the Python-based workflow, DBOS middleware extracts source and destination IP addresses from each HTTP request and appends them to in-memory vectors. As shown in Figure~\ref{fig:local_python}, a base GraphBLAS traffic matrix is constructed from the source-destination vectors using Python-GraphBLAS every $N_v$ requests and serialized to disk. Once $N_b$ new base matrices are created, they are deserialized and hierarchically summed to form a locally aggregated GraphBLAS matrix. This traffic matrix is saved and the nine network quantities are computed on the traffic matrix  before being sent to the coordinator.

\subsection{Local Aggregation (PostgreSQL)}
\begin{figure}[htbp]
\centering
\includegraphics[width=1.0\columnwidth]{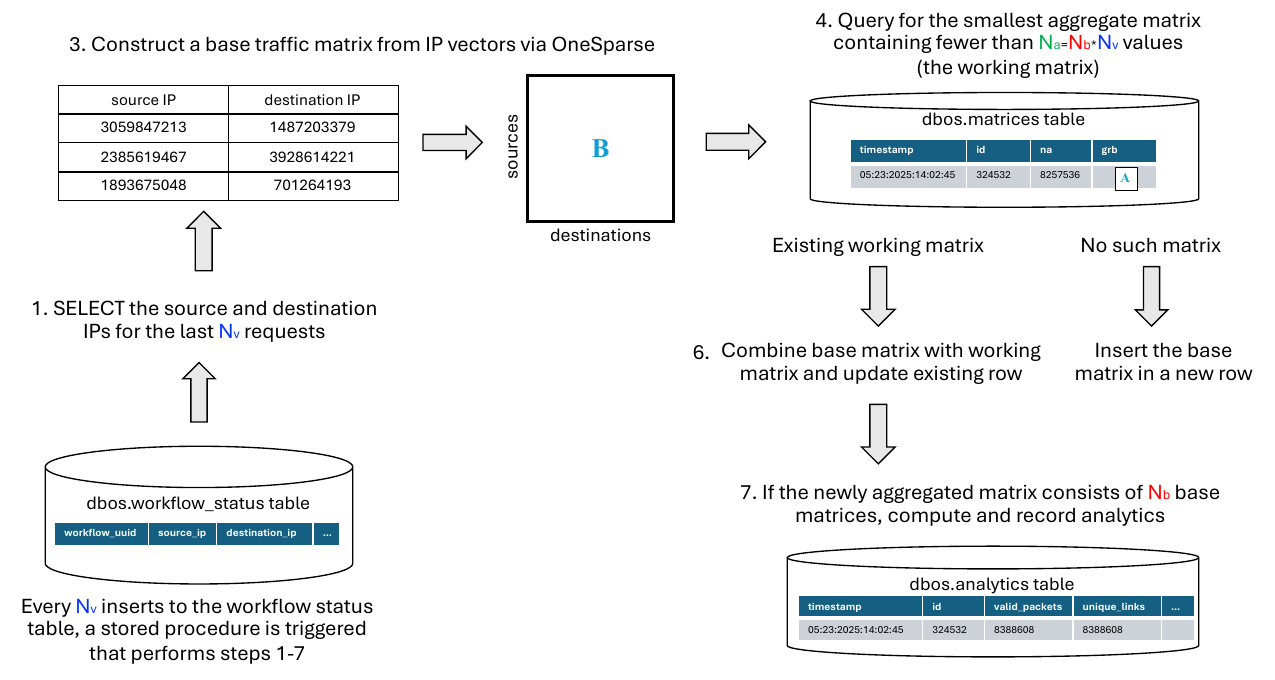}
\caption{\textbf{PostgreSQL Local Aggregation Workflow}}
\label{fig:postgresql}
\end{figure}

The PostgreSQL workflow mirrors the Python variant but relies on SQL triggers and stored procedures. As shown in Figure~\ref{fig:postgresql}, a trigger on the \texttt{dbos.workflow\_status} table fires every insert, incrementing a PostgreSQL sequence that serves as a counter. When the counter reaches a multiple of $N_v$, the latest $N_v$ IP pairs are selected from the \texttt{dbos.workflow\_status} table to construct a base matrix using OneSparse. Instead of storing each base matrix as a separate row for later batch processing, each new matrix is incrementally added to an in-place working aggregate. Once $N_b$ base matrices are combined and the working aggregate reaches its final request count of $N_a = N_v*N_b$ values, summary analytics are computed and stored in the \texttt{dbos.analytics} table. The next base matrix produced is inserted as a new entry in the \texttt{dbos.matrices} table to initialize a new working aggregate matrix and begin the next aggregation cycle.

\subsection{Global Aggregation}
\begin{figure*}[htbp]
\centering
\includegraphics[width=0.9\textwidth]{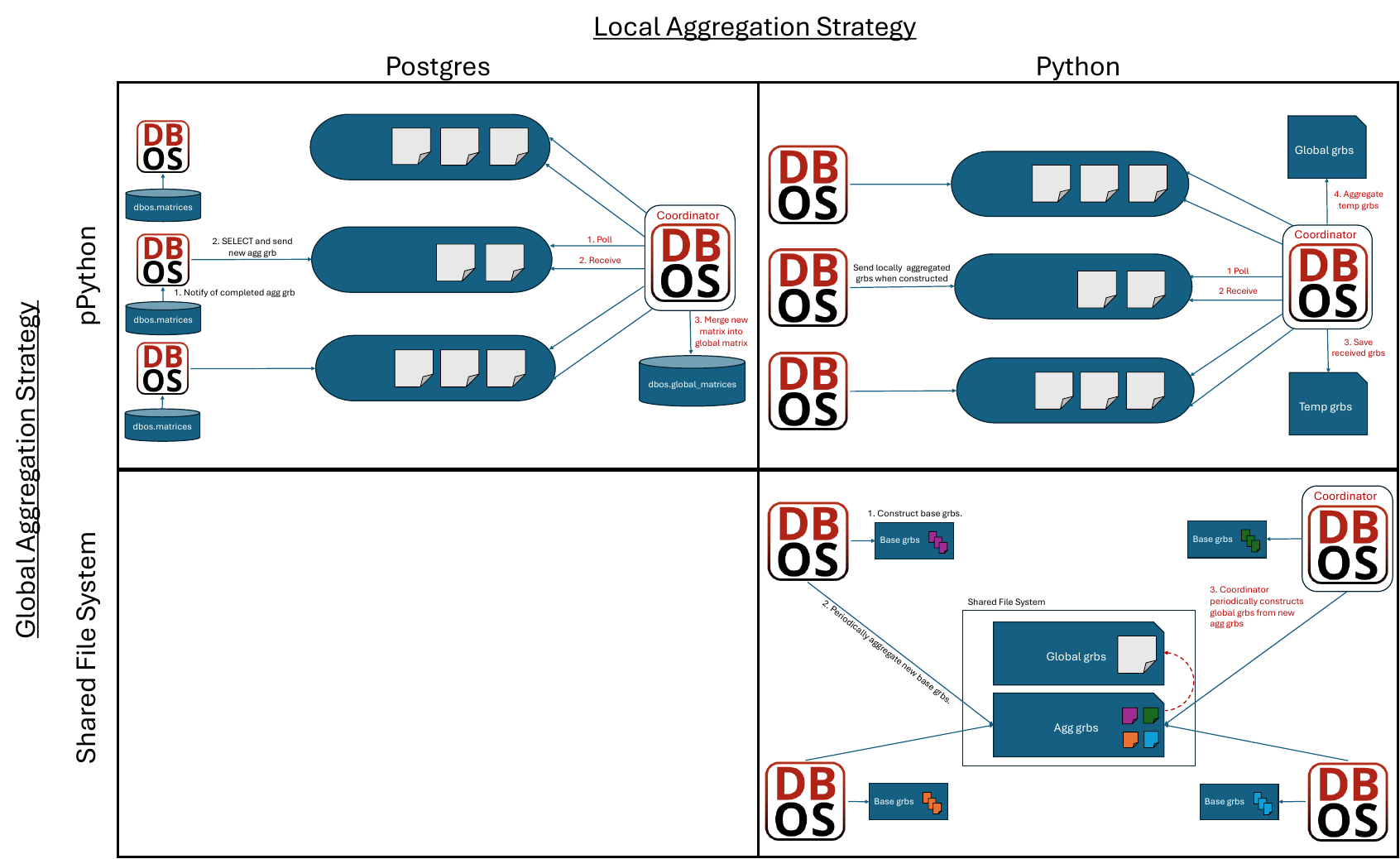}
\caption{\textbf{Comparison of Aggregation Strategies}. Summary of different coordination mechanisms (pPython message passing vs. shared file system (SFS)) across local workflows (PostgreSQL vs. Python). Each cell illustrates the workflow steps and data flow between worker nodes and the coordinator.}
\label{fig:Aggregation2x2}
\end{figure*}

DBOS supports two global aggregation strategies: message passing with pPython and coordination via a shared file system. As illustrated in Figure~\ref{fig:Aggregation2x2}, both follow a centralized design in which locally aggregated matrices are transmitted to a coordinator node for final aggregation and analysis.

The message-passing approach uses pPython, a simple, file-based implementation of the Message Passing Interface (MPI) standard in Python, designed to operate entirely within user space without escalated privileges \cite{byun2022ppython, byun2023ppython}. Each process is assigned a rank: rank 0 serves as the coordinator, while all others act as workers. Communication is performed with three primitives: \texttt{SendMsg} to transmit serialized matrices, \texttt{ProbeMsg} to check for incoming messages, and \texttt{RecvMsg} to retrieve them.

In Python-based workflows, each worker invokes \texttt{SendMsg} immediately after generating a locally aggregated matrix. PostgreSQL workers, however, cannot call \texttt{SendMsg} directly from within the database. Instead, they issue a notification on a PostgreSQL channel when a new aggregate is ready. A companion Python process listens for these notifications, which include the ID of the completed matrix. Upon receiving a notification, the process queries the \texttt{dbos.matrices} table for the corresponding matrix and forwards it to the coordinator using \texttt{SendMsg}.

The coordinator continuously probes for incoming messages and processes each received matrix using \texttt{RecvMsg}. In PostgreSQL-based workflows, each matrix is incrementally merged into a working global matrix stored in the \texttt{dbos.global\_matrices} table, following the same accumulation pattern used for PostgreSQL local aggregation. In Python-based workflows, the coordinator writes incoming matrices to temporary files and performs hierarchical aggregation once $N_g$ matrices have been received, also replicating the same aggregation strategy used for local Python aggregation.

\section{Results}
\subsection{Experimental Setup}
All experiments were conducted on MIT SuperCloud nodes each with 48 2.4 GHz x86 processing cores, 192 GB of RAM, with a 25 Gbit network connection to a  Lustre shared file system (SFS) \cite{reuther2018interactive}. Each job was launched using pPython with a single coordinator on a node and a configurable number of workers, each running on their own separate node. Workers processed synthetic IP pair data at a constant injection rate of $10^5$ requests per second (RPS). This rate is approximately one order of magnitude lower than the peak throughput sustained by the PostgreSQL workflow (the lowest throughput workflow), and one order of magnitude higher than the expected maximum request rate of a real-world DBOS deployment.

To stress-test the system beyond typical web server workloads, we bypassed portions of the standard DBOS request pipeline. In the Python workflows, we generated synthetic requests by creating randomized NumPy vectors of $N_v$ 32-bit unsigned integers. These vectors were passed into the aggregation workflow in place of individual IP pairs. For the PostgreSQL workflow, per-request inserts into the \texttt{dbos.workflow\_status} table proved prohibitively slow due to the transactional overhead incurred by each insert. These inserts are part of the standard DBOS request processing pipeline, and thus impose a practical upper bound on the system’s sustainable throughput under real workloads. To mitigate this, we constructed base matrices using Python-GraphBLAS and inserted them directly into the matrices table.

We adopted the Graph Challenge’s recommended value of $N_v = 2^{17}$ web requests to form into each base matrix. Based on benchmarks for throughput and memory overhead for different matrix densities, to maximize aggregation efficiency without exceeding system memory limits, we configured the number web requests in each local aggregate matrix as $N_a = 2^{23}$ and each global aggregate matrix as $N_g = 2^{25}$.

\subsection{Resource Utilization}
\begin{figure*}
\centering
\includegraphics[width=0.9\textwidth]{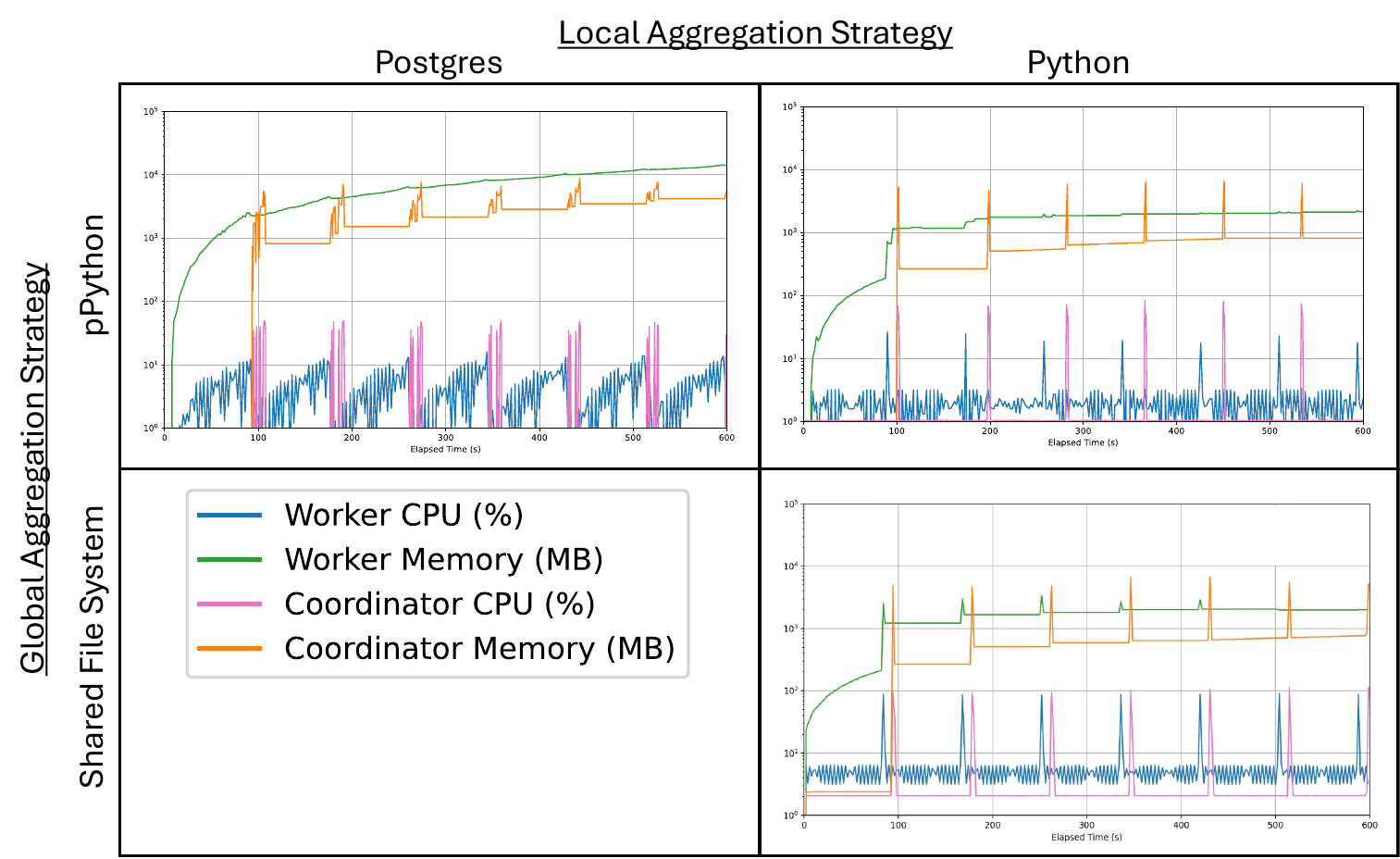}
\caption{\textbf{Representative Example Resource Utilization.} CPU and memory usage for each local-global strategy pair for a run with 1 coordinator and 4 workers. Worker metrics are medians across all 4 workers. Resource statistics were recorded by running Python psrecord with a probing interval of 0.5 seconds over a 10-minute benchmark window. CPU utilization is expressed as a percentage of the 48-core capacity.}
\label{fig:resource_utilization}
\end{figure*}

Figure~\ref{fig:resource_utilization} displays representative CPU and memory usage for each combination of local and global aggregation strategies. Resource utilization patterns align closely with workflow stages: spikes in coordinator CPU and memory usage correspond to global aggregations and analyses, while spikes in median worker resource usage reflect local aggregations and analyses. In the recorded experiment, each global aggregation is triggered after four local aggregations ($N_g = 4N_a$) across four workers, resulting in a one-to-one correspondence between global and local aggregation spikes.

\begin{figure*}[htbp]
\centering
\includegraphics[width=0.9\textwidth]{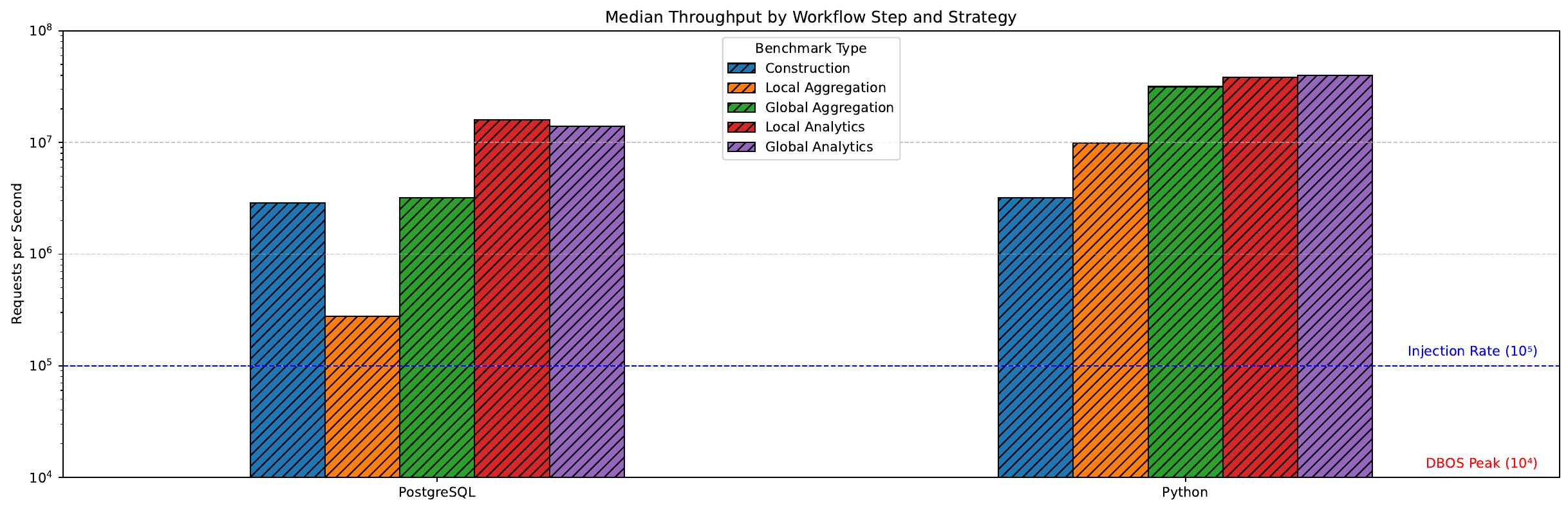}
\caption{\textbf{Single Worker Single Coordinator Median Throughput of Each Workflow Stage.} Each bar shows the median requests per second (RPS) achieved during a 10-minute interval for matrix construction, aggregation, and analytics.}
\label{fig:benchmark_medians}
\end{figure*}

Worker resource usage is fairly consistent across time and strategies, with the PostgreSQL strategy proving to be the most resource-intensive. Both Python workflows (pPython and SFS) exhibit similar and relatively stable resource profiles. Aggregation triggers brief spikes in CPU and memory utilization. For the PostgreSQL workflow, this creates a characteristic sawtooth pattern, reflecting the increasing cost of summing progressively larger working aggregate matrices.

Across all workflows, CPU usage remains low for most stages and approaches 100\% only during the aggregation and analysis phase. PostgreSQL’s CPU bottlenecks stem from its single-threaded design. Each worker process handles one connection, and matrix operations remain single-threaded unless explicitly offloaded. During aggregation and analysis, we invoke the GraphBLAS backend via OneSparse, which supports multithreaded operations. This is the only phase where PostgreSQL achieves meaningful parallelism. The Python workflows also suffer from limited parallelism due to Python's Global Interpreter Lock (GIL), which prevents concurrent execution of Python bytecode. However, the GraphBLAS backend used in aggregation is implemented in C and releases the GIL, enabling multithreaded execution. As a result, we only observe significant spikes in CPU usage during aggregation and analysis.

Overall, the system is rate-limited by the sequential, bursty, and CPU-bound workloads caused by periodic aggregation. Despite the single-threaded limitations, resource usage remains reasonable for all strategies at an injection rate much higher than practically necessary.

\subsection{Workflow Throughput}
Figure~\ref{fig:benchmark_medians} reports the median throughput achieved at each stage of the pipeline for the two workflow types. Python-based workflows consistently outperform the PostgreSQL implementation across all stages. Although both rely on the same underlying GraphBLAS backend, PostgreSQL incurs significantly more overhead due to transaction management, PostgreSQL serialization overhead, and repeated compression and decompression of matrix objects.

To contextualize system performance, we loosely compare against the Graph Challenge reference benchmarks (see Figure 7 in \cite{jananthan2024anonymized}). Our implementation achieves throughput of $10^{6.2}$ RPS in construction, $10^7$ RPS in aggregation, and $10^{7.3}$ RPS in analytics using a single process with
48 threads. Reported throughput for the reference implementation is approximately $10^7$ packets
per second for construction, $10^{7.7}$ for aggregation, and $10^{7.2}$. The reference implementation uses 48 parallel construction processes and three processes for aggregation and analysis with 48 core nodes. Given that our workflows execute sequentially within a single process, performance remains in the same regime as the reference benchmarks.

In summary, the PostgreSQL workflow provides durability and transactional consistency, but at the cost of throughput and resource efficiency. Python-based workflows, while lacking the same degree of persistence, are significantly more performant and better suited to high-throughput workloads.

\subsection{Scalability}
The scalability of each workflow is fundamentally constrained by the coordinator's ability to process and aggregate incoming matrices. The system remains scalable only if the coordinator can complete global aggregation and analysis before the next batch of locally aggregated matrices arrives. When this condition is violated, the system begins to fall behind, leading to performance degradation.

Global aggregation is triggered once $N_g/N_a$ locally aggregated matrices have been produced. Benchmark results show that global aggregation and analysis takes a median of $1.9$ seconds for Python workflows and $12.9$ seconds for PostgreSQL workflows, yielding effective coordinator throughputs of $10^{7.2}$ and $10^{6.4}$ RPS, respectively.

\begin{figure}[htbp]
\centering
\includegraphics[width=1.0\columnwidth]{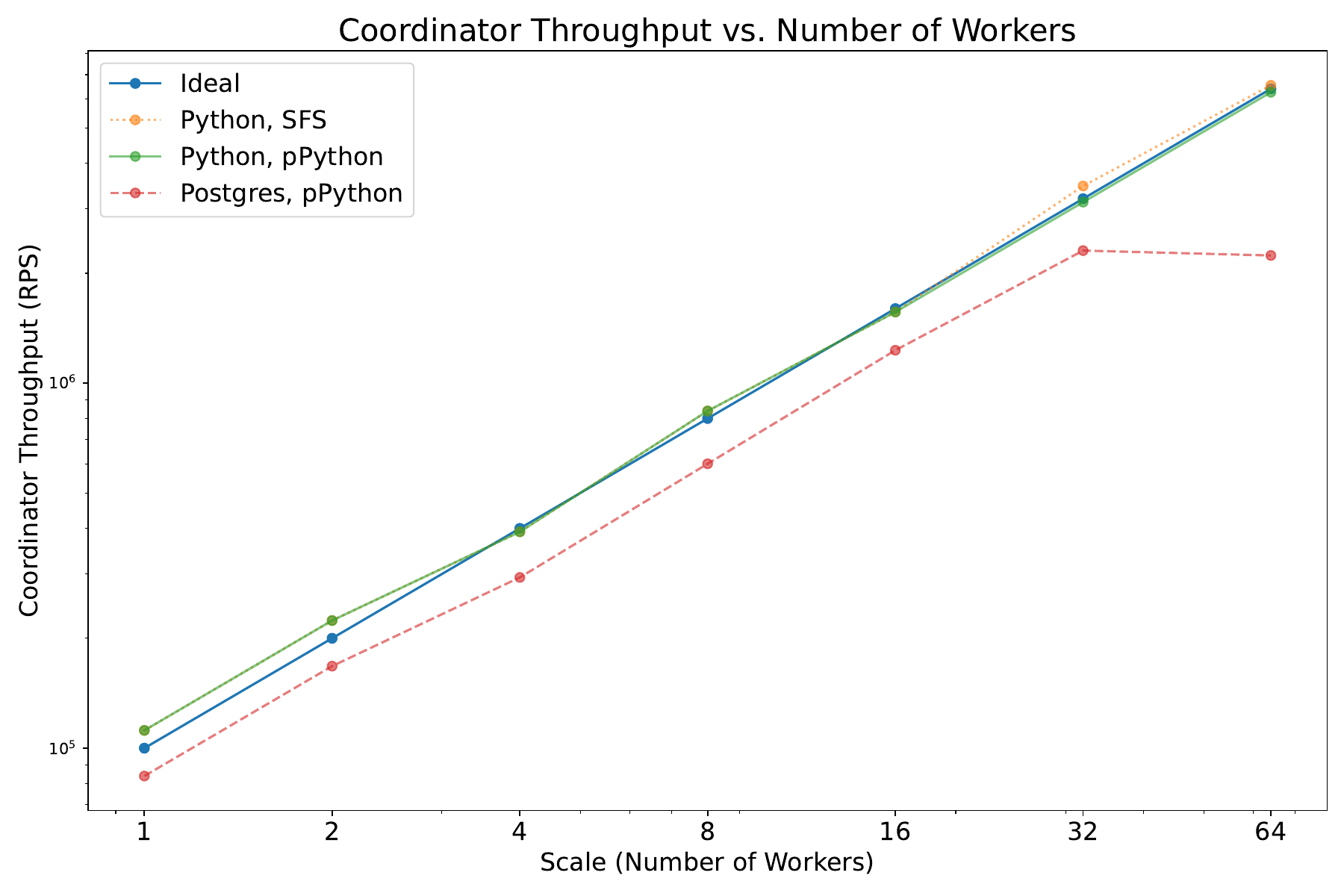}
\caption{\textbf{Coordinator Throughput vs Worker Count.} Coordinator throughput is measured by the number of global matrices produced in a 10 minute window. Each global matrix corresponds to $N_g=2^{25}$ requests so throughput is calculated as $\frac{G*N_g}{600}$ where $G$ is the number of global matrices produced by the coordinator. The coordinator throughput should scale linearly with the number of workers.}
\label{fig:matrix_scaling}
\end{figure}

At a constant injection rate of $10^5$ RPS per worker, these throughput limits correspond to a theoretical maximum of ${\approx}175$ Python workers or ${\approx}25$ PostgreSQL workers. If the injection rate is reduced to DBOS’s typical peak of $10^4$ RPS, the system could support ${\approx}1750$ Python-based workers or ${\approx}250$ PostgreSQL-based workers. These limits are consistent with empirical results, shown in Figure~\ref{fig:matrix_scaling}, where Python-based workflows scale linearly in coordinator request throughput up to $64$ workers. In contrast, the PostgreSQL workflow plateaus at a maximum throughput of about $10^{6.35}$ RPS over a $10$ minute period. This is slightly below the theoretical maximum due to CPU contention with the Python process responsible for probing and receiving incoming matrices. However, the PostgreSQL workflow still closely aligns with the estimated processing limit of the coordinator and throughput scales linearly up to 32 workers.

\begin{figure*}
\centering
\includegraphics[width=0.9\textwidth]{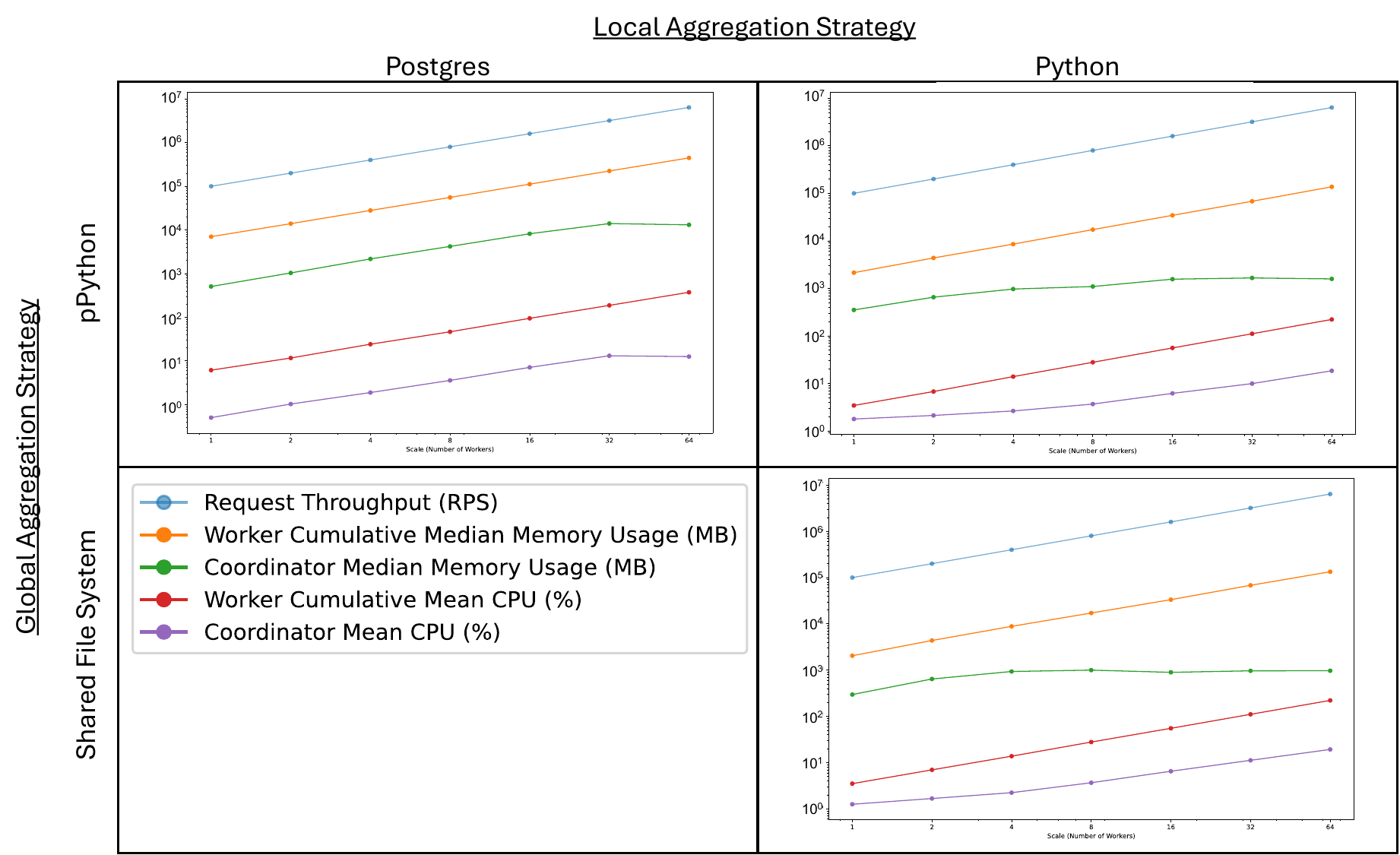}
\caption{\textbf{Resource Usage vs. Worker Count Comparison.} Worker CPU and memory usage are summed across all workers. Coordinator values reflect mean CPU utilization and median memory usage over a 10-minute window.}
\label{fig:sfs_resource_scale}
\end{figure*}

Coordinator resource usage is consistent with the worker measurements. As shown in Figure~\ref{fig:sfs_resource_scale}, coordinator CPU and memory usage scales linearly in the PostgreSQL workflow, but remains sublinear for Python workflows. This is attributed to two factors: (1) aggregation occurs in short bursts, often underrepresented by the 0.5-second sampling interval, and (2) memory is explicitly garbage collected after each aggregation, allowing reuse across batches.

\section{Conclusion}

This work introduces an implementation of collaborative network awareness embedded directly into a web services runtime. By integrating the GraphBLAS traffic matrix framework within DBOS, we demonstrate that network traffic matrix analytics can be performed in real time as a native web service capability, without reliance on external agents or third-party monitoring infrastructure.  We present and evaluate two local aggregation workflows (Python and PostgreSQL) and two coordination mechanisms for global aggregation (message passing via pPython and a shared file system). Experimental results show that both Python-based workflows consistently exceed the DBOS maximum expected RPS. Our evaluation confirms that the system scales linearly with the number of workers up to the point at which the coordinator becomes the performance bottleneck. At least up to 64 workers, the strategies track closely to theoretical scaling limits.

While both workflows are single-threaded in design, the architecture could be modified to achieve multi-process parallelism. A stream of network events arriving at a single DBOS instance can be partitioned into multiple sub-streams, each handled by an independent process executing the same workflow logic on its assigned portion of the data. This offers a clear path to higher throughput on multi-core machines in high-performance environments.

In summary, this paper establishes a practical and performant foundation for embedding network analysis into the DBOS web service. By shifting collaborative awareness from an external service to an intrinsic capability of the web service, DBOS with GraphBLAS traffic matrices offers a powerful model for building resilient, secure, and high-throughput systems. Future extensions, including integrated anonymization, real-world traffic deployment, coordinated response mechanisms, and long-term historical data management, could further advance this platform towards providing integrated collaborative awareness. 

\section*{Acknowledgments}

The authors wish to acknowledge the following individuals for their contributions and support:  Sean Atkins, Chris Birardi, Bob Bond, Bill Cashman, K  Claffy, Chris Demchak, Alan Edelman, Jeff Gottschalk,  Daniel Grant,  Thomas Hardjono, Chris Hill, Charles Leiserson, Sam Madden, Kristen Malvey, Sanjeev Mohindra, Heidi Perry, Sandeep Pisharody, Christian Prothmann,  Steve Rejto, Scott Ruppel, Daniela Rus,  Mark Sherman, Shahar Somin, Scott Weed,  Marc Zissman.

\bibliographystyle{ieeetr}
\bibliography{DBOS-AnonNetSensing}

\end{document}